# Super-stretchable borophene and its stability under straining

Zhenqian Pang[a], Xin Qian[b], Ronggui Yang[b*], Yujie Wei[a*]

[a]LNM, Institute of Mechanics, Chinese Academy of Sciences, Beijing, 100190, People's Republic of China

[b]Department of Mechanical Engineering and Materials Science and Engineering Program, University of Colorado, Boulder, CO 80309, USA

*Corresponding authors: (Y. Wei) yujie_wei@lnm.imech.ac.cn; (R.G. Yang) ronggui.yang@colorado.edu

Recent success in synthesizing two-dimensional borophene on silver substrate attracts strong interest in exploring its possible extraordinary physical properties. By using the density functional theory calculations, we show that borophene is highly stretchable along the transverse direction. The strain-to-failure in the transverse direction is nearly twice as that along the longitudinal direction. The straining induced flattening and subsequent stretch of the flat borophene are accounted for the large strain-to-failure for tension in the transverse direction. The mechanical properties in the other two directions exhibit strong anisotropy. Phonon dispersions of the strained borophene monolayers suggest that negative frequencies are presented, which indicates the instability of free-standing borophene even under high tensile stress.

Since the successful synthesis of single-layer hexagonal boron-nitride (h-BN)[1, 2], the community of two-dimensional materials has been curious about the possibility of making monolayer borophene [3-5]. Boron has similar valence orbitals residing in the left side of carbon in the periodic table, it has the potential to form new 2-D structure to extend the nanostructured materials [6-8]. Nevertheless, two dimensional boron cannot form stable honeycomb hexagonal structure [3] like graphene.[9-11] Triangular boron lattices with hexagonal vacancies [7] were predicted to be stable. Liu et al showed the growth mechanism and hole formation from boron cluster to 2-D boron sheet on Cu (1 1 1) surface by first-principles calculations. Now B36 with a central hexagonal hole has been seen in experiments.[9] A more flexible structure for borophene, such as B35 cluster with a double-hexagonal vacancy[12] and B30 [13], are also proposed by DFT calculations. Until very recently, monolayer borophene on silver substrate has been observed by scanning tunneling spectroscopy (STS).[14] It is of great interests to examine the physical properties that might be originated from its two-dimensional nature and to see such two-dimensional structure can be free-standing with mechanical stability. Typically, a buckled status of monolayers could endure more stretch from the macroscopic point of view.[15] In this work, we perform a systematic investigation on the mechanical properties of borophene, including the Young's modulus, the Poisson ratio, the strength, and the phonon dispersions of the strained borophene monolayers.

First principles density functional theory (DFT) calculations on borophene were performed with the Vienna Ab initio Simulation Package (VASP) code[16, 17] The projector augmented wave (PAW) pseudopotentials[18, 19] and the generalized gradient approximation (GGA) of the Perdew-Burke-Ernzerhof (PBE) functional[20, 21] are used. A plane-wave basis set with a kinetic-energy cut-off of 400 eV and a Monkhorst-Pack[22] k-point mesh of 31×31×1 are used for static electronic structure calculations. To eliminate the interactions between periodic images of borophene a vacuum space of 20Å was used. Periodic boundary conditions are applied to the two in-plane directions in all the calculations conducted here. All structures are relaxed using a conjugate gradient algorithm until the atomic forces are converged to 0.01eV/Å.

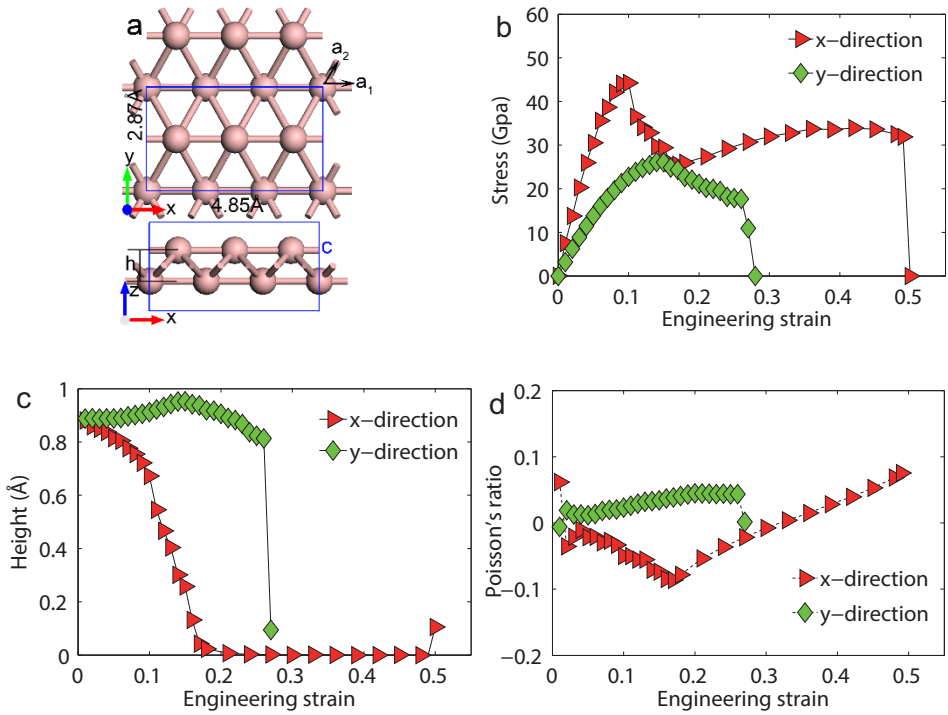

Figure 1. (Color online) Mechanical behavior of monolayer borophene under uniaxial tension: (a) The monolayer viewed from the top and the side. (b) The stress-strain curves for uniaxial loading along the longitudinal (x-direction) and the transverse direction (y-direction). (c) Evolution of the layer height under straining. (d) The Poisson's ratio versus strains.

As illustrated in Figure 1a, the orientation of the single-layer crystalline borophene is described by two vectors $a_1$ and $a_2$ ,. Mechanical loading is applied to two different orientations, (I) transverse $a$ =(1,0); (II) longitudinal $a$=(2,1). Included in Figure 1a is the equilibrium lattice constants obtained by DFT calculations, which matches well with the experimental measurement of 5.10 Å by 2.9 Å. [14] When calculating the equivalent stress of borophene, we assume an interlayer distance $c$=4.8Å . We first show the engineering stress versus engineering strain curves for the samples subjected to uniaxial tension in the transverse and longitudinal directions (Figure 1b). Similar to the mechanical properties of h-BN [26], we find that the strength and the failure strain in borophene are strongly anisotropic. During tension in the transverse direction, the material shows higher strength and greater failure strain. There exist two stress peaks in the stress-strain curves compared with that in the longitudinal direction, the first occurs at strain of 10% and the second appears at a strain of 45%. From Fig. 1c, we see that the layer height decreases with the increasing strain and the structure eventually flattens out with h = 0. Interestingly, the transverse stress increases again when the sample is fully flattened (at 18% strain); under tension in longitudinal direction, the samples fails at strain of 27%. Figure 1d shows the Poisson's ratio $v$ as a function of strain. The Poisson's ratio is defined as $v = -\frac{L_L - L0_L}{L_T - L0_T}\frac{L0_T}{L0_L}$, where $L0_L$ and $L0_T$ are the initial lengths of the unit cell in the loading direction and perpendicular direction, respectively, while $L_L$ and $L_T$ are the corresponding ones after tension. When tension is applied in the transverse direction, the Poisson's ratio is negative which indicates resistance of load is positive in the other direction. Similar to the single-layer black phosphorus [27] and penta-graphene[28], the negative Poisson's ratio originates from the buckled structure.

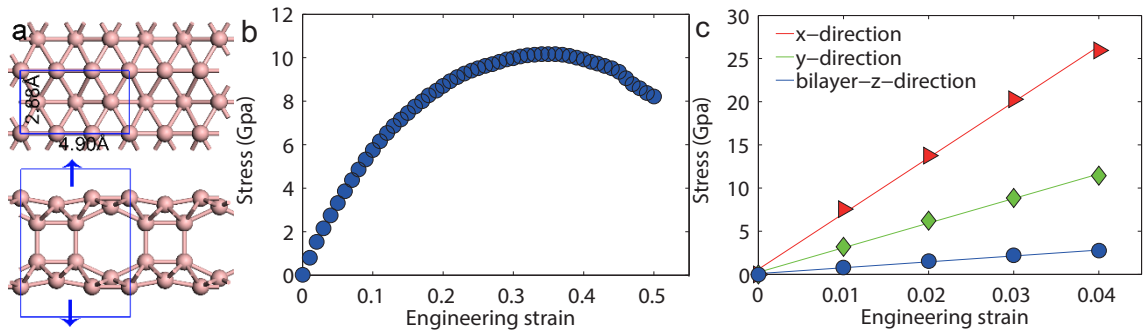

Figure 2. (Color online) Anistropic elastic properties of borophene. (a) The bilayer borophene viewed from the top and the side. (b) The stress as a function of the strain in z-direction. (c) The stress-strain curves under small elastic strain for tensile loading along three directions.

Figure 2a shows the structure and lattice constants of a bilayer borophene. Different from other layered structures [29] where interlayer interaction is of Van der Waals bond, there are chemical bonds between the two layers of bilayer borophene.  By applying stretch perpendicular to boron plane to tear apart the two layers, we see that the peak stress occurs at a strain of 38%. At the same time, the separation strength could be as high as 10GPa, which is considerably greater than that of graphene (about 3GPa). In Fig. 2c, we show the detailed calculations for the two in-plane Young's modulus in monolayer borophene and the out-of-plane Young's modulus in the bilayer borophene. The Young's modulus in the x (transverse) direction is about 646.6Gpa, that in the y-direction is 285.2Gpa, and the modulus along the z-direction is the smallest, about 68.8Gpa.

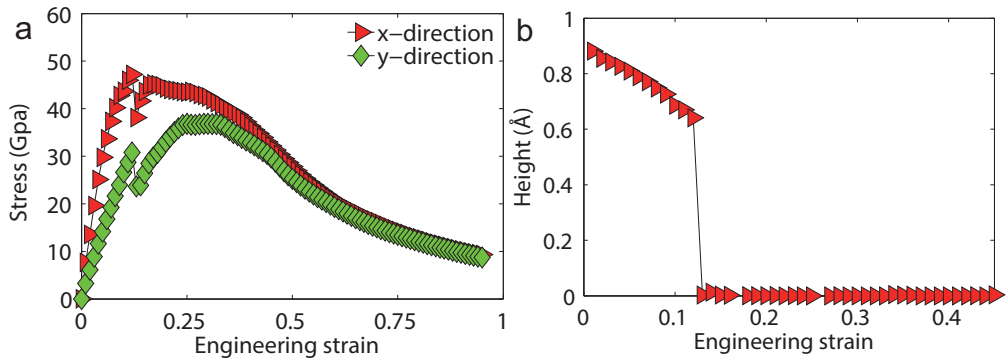

Figure 3. (Color online) The mechanics of monolayer borophene under biaxial tension. (a) The stress as a function of strain in both directions. (b) The height as a function of the biaxial strain.

As a comparison, we also explore the stress-strain response of borophene under biaxial tension as shown in Figure 3a. A unique mechanic response is observed during the biaxial stretching. A sudden stress drop occurs at 12 % strain in the two stress-strain response curves, which is due to the flattening of the borophene from the buckled structure to the planar structure. This structure transition, however, happens gradually as we apply uniaxial stretch along the x direction. After the sudden stress drop, the stress increases again and the second peak appears. We relax the structures for strains within 18%±8% by assigning a small perturbation. Structures with strain larger than 17% will evolve to a disorder state. It indicates that the maximum biaxial strain is 17%; beyond this point, the structure is not stable from the total energy point of view.

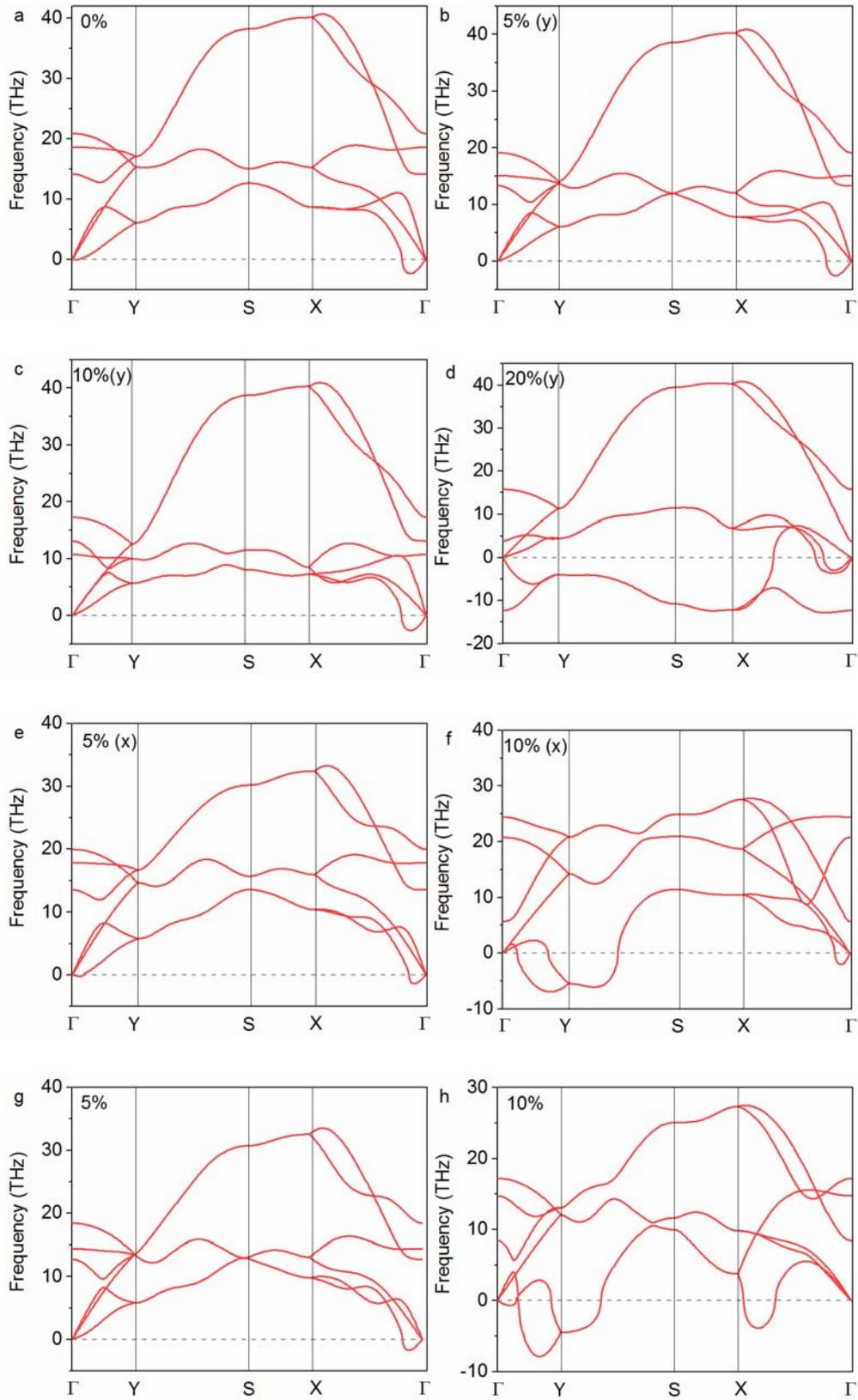

Figure 4. (Color online) Phonon dispersions of the strained borophene monolayers. (a) The phonon dispersion at equilibrium. (b) to (d) The phonon dispersion at under y-axis (a=(2,1)) strain of 5%, 10% and 20%, respectively. (e)

to (f) The Phonon dispersion at the x-axis (a=(1,0))strain of 5% and 10%, in turn. (g) and (h) The Phonon dispersion at the bi-axial strain of 5% and 10%, respectively.

To further examine the stability of stretched borophene, we calculate the phonon dispersions of the strained borophene monolayers using the finite displacement method [23, 24], implemented in the Phonopy [25] package. Small displacements (<0.06 Å) are first imposed on the inequivalent B atoms in a 2x4x1 supercell. With the perturbed structures, DFT calculations are then conducted to extract the interatomic forces as a response to the corresponding displacement. Phonon dispersions are calculated by Phonopy package[25] based on the relationship between interatomic forces and atomic displacements. To make sure that the negative frequencies of ZA mode are not caused by the selection of k-mesh density[30], we performed various calculations based on k-mesh sizes of 4x4x1, 12x12x1 and 16x16x1.

Figure 4 shows the phonon dispersions of the strained monolayer borophene generated using a k-mesh density of 16x16x1. We have applied both isotropic biaxial strains and uniaxial strains along x and y axes For the phonon dispersion of the initial unstrained structure (Figure 4a), negative frequencies are observed in the out-of-plane acoustic mode (ZA mode) near Brillouin zone centre along $\Gamma - X$ and $\Gamma - Y$ directions. Negative frequencies mean that the force constants of these vibration modes are negative, therefore restoring forces cannot be generated for these modes. The structure tends to go away from its original configuration. For small strains (5%) applied (Figure 4b, e and g), the negative frequency of the ZA branch in $\Gamma - Y$ direction is eliminated, but is still present in $\Gamma - X$ direction. For larger strains applied (Figure 4c, d, f and h), the in-plane transverse (TA) modes are also destabilized.

We therefore conclude that the freestanding borophene is mechanically unstable even with strains, and the stability of synthesized borophene is provided by the out-of-plane constraints from the silver substrates.

In summary, we investigated several critical mechanical properties of monolayer and bilayer borophene by using first-principles DFT calculations. We find that the failure behavior, Young's modulus and the Poisson's ratio of borophene are highly anisotropic: Poisson's ratio is negative in tension along transverse direction. Furthermore, when stretching along z-direction in bilayer borophene, the strength and failure strain are much higher than that of bilayer graphene. In addition, we investigated the mechanics of monolayer borophene under biaxial tension and we found that biaxial tension increases the strength in longitudinal direction and decreases the failure strain in transverse direction. Calculations for phonon dispersion suggest that the instability of out-of-plane modes are present for free-standing borophene under high level tensile stress. Our phonon dispersion calculations suggest that the stability of synthesized borophene might be provided by constraints of the out-of-plane vibrations from the silver substrates..

**Acknowledgements**

Y.W. acknowledges the support from the National Natural Science Foundation of China (NSFC) (Grant no. 11425211), and MOST 973 of China (Grant no. 2012CB937500). R.Y. acknowledges the support from National Science Foundation (Grant No. 1512776). Calculations of Mechanical Properties are performed at the Supercomputing Center of CAS. Calculations of Phonon Dispersion are performed using Janus supercomputer supported by National Science Foundation (Grant No.0821794) and University of Colorado Boulder.

**References**

1. Novoselov, K.S., et al., *Two-dimensional atomic crystals.* Proc Natl Acad Sci U S A, 2005. **102**(30): p. 10451-10453.
2. Chopra, N.G., et al., *Boron Nitride Nanotubes.* Science, 1995. **269**(5226): p. 966-967.
3. Lau, K.C. and R. Pandey, *Stability and electronic properties of atomistically-engineered 2D boron sheets.* The Journal of Physical Chemistry C, 2007. **111**(7): p. 2906-2912.
4. Penev, E.S., et al., *Polymorphism of two-dimensional boron.* Nano letters, 2012. **12**(5): p. 2441-2445.
5. Liu, Y., E.S. Penev, and B.I. Yakobson, *Probing the Synthesis of Two-Dimensional Boron by First-Principles Computations.* Angewandte Chemie International Edition, 2013. **52**(11): p. 3156-3159.
6. Zhai, H.-J., et al., *Observation of an all-boron fullerene.* Nat Chem, 2014. **6**(8): p. 727-731.
7. Yang, X., Y. Ding, and J. Ni, \textit{Ab initio} *prediction of stable boron sheets and boron nanotubes: Structure, stability, and electronic properties.* Physical Review B, 2008. **77**(4): p. 041402.
8. Tang, H. and S. Ismail-Beigi, *Novel precursors for boron nanotubes: the competition of two-center and three-center bonding in boron sheets.* Physical review letters, 2007. **99**(11): p. 115501.
9. Piazza, Z.A., et al., *Planar hexagonal B36 as a potential basis for extended single-atom layer boron sheets.* Nat Commun, 2014. **5**.
10. Lau, K.C. and R. Pandey, *Stability and Electronic Properties of Atomistically-Engineered 2D Boron Sheets.* Journal of Physical Chemistry C, 2007. **111**: p. 2906-2912.
11. Evans, M.H., J.D. Joannopoulos, and S.T. Pantelides, *Electronic and mechanical properties of planar and tubular boron structures.* Physical Review B, 2005. **72**(4).
12. Li, W.-L., et al., *The B35 Cluster with a Double-Hexagonal Vacancy: A New and More Flexible Structural Motif for Borophene.* Journal of the American Chemical Society, 2014. **136**(35): p. 12257-12260.
13. Li, W.-L., et al., *[B30]−: A Quasiplanar Chiral Boron Cluster.* Angewandte Chemie, 2014. **126**(22): p. 5646-5651.
14. Mannix, A.J., et al., *Synthesis of borophenes: Anisotropic, two-dimensional boron polymorphs.* Science, 2015. **350**(6267): p. 1513-1516.
15. Wang, B., et al., *Stable planar single-layer hexagonal silicene under tensile strain and its anomalous Poisson's ratio.* Applied Physics Letters, 2014. **104**(8): p. 081902.
16. Kresse, G. and J. Furthmüller, *Efficiency of ab-initio total energy calculations for metals and semiconductors using a plane-wave basis set.* Computational Materials Science, 1996. **6**(1): p. 15-50.
17. Kresse, G. and J. Furthmüller, *Efficient iterative schemes for ab initio total-energy calculations using a plane-wave basis set.* Physical Review B, 1996. **54**(16): p. 11169-11186.
18. Blöchl, P.E., *Projector augmented-wave method.* Physical Review B, 1994. **50**(24): p. 17953-17979.

19. Kresse, G. and D. Joubert, *From ultrasoft pseudopotentials to the projector augmented-wave method.* Physical Review B, 1999. **59**(3): p. 1758-1775.
20. Perdew, J.P., K. Burke, and M. Ernzerhof, *Generalized Gradient Approximation Made Simple.* Physical Review Letters, 1996. **77**(18): p. 3865-3868.
21. Perdew, J.P., K. Burke, and M. Ernzerhof, *Generalized Gradient Approximation Made Simple [Phys. Rev. Lett. 77, 3865 (1996)].* Physical Review Letters, 1997. **78**(7): p. 1396-1396.
22. Monkhorst, H.J. and J.D. Pack, *Special points for Brillouin-zone integrations.* Physical Review B, 1976. **13**(12): p. 5188-5192.
23. Parlinski, K., Z.Q. Li, and Y. Kawazoe, *First-principles determination of soft mode in cubic ZrO2.* Physical Review Letters, 1997. **78**(21).
24. Kresse, G., J. Furthmuller, and J. Hafner, *Ab initio Force Constant Approach to Phonon Dispersion Relations of Diamond and Graphite.* Europhysics Letters, 1995. **32**(9): p. 729-734.
25. Togo, A. and I. Tanaka, *First principles phonon calculations in materials science.* Scripta Materialia, 2015. **108**: p. 1-5.
26. Wu, J., et al., *Mechanics and Tunable Bandgap by Straining in Single-Layer Hexagonal Boron-Nitride.* arXiv preprint arXiv:1301.2104, 2013.
27. Jiang, J.-W. and H.S. Park, *Negative poisson's ratio in single-layer black phosphorus.* Nat Commun, 2014. **5**.
28. Zhang, S., et al., *Penta-graphene: A new carbon allotrope.* Proceedings of the National Academy of Sciences, 2015. **112**(8): p. 2372-2377.
29. Rydberg, H., et al., *Van der Waals Density Functional for Layered Structures.* Physical Review Letters, 2003. **91**(12): p. 126402.
30. Şahin, H., et al., *Monolayer honeycomb structures of group-IV elements and III-V binary compounds: First-principles calculations.* Physical Review B, 2009. **80**(15).